# The Least Action-Augmented Lanchester Model

Wei Liang, Han Hu, Lijie Sun, Pingxing Chen and Ming Zhong[*]

College of Science, National University of Defense Technology, Changsha 410073, People's Republic of China

[*]E-mail: zhongm@nudt.edu.cn

**Abstract**

The principle of least action, a fundamental principle in variational mechanics with broad applicability to classical physical systems, is employed to formulate a novel attrition model for combat dynamics. This formulation extends the Lanchester's square law through second-order temporal derivatives by requiring the resultant Euler-Lagrange equation to coincide with the classical Lanchester's equation. Initial conditions at a specified temporal point enable determination of subsequent system evolution through action minimization, while terminal boundary conditions permit backward reconstruction of combat trajectories. The model's validity is examined through historical analysis of WWII engagements: the Battle of Kursk and the Battle of Iwo Jima. Comparative studies with conventional Lanchester's square models demonstrate marked improvements in predictive accuracy regarding force strength progression, particularly in capturing non-linear attrition patterns characteristic of prolonged engagements.

# 1 Introduction

Lanchester-type models (Lanchester, 1916) have demonstrated extensive applicability in analyzing competitive resource-depletion processes across multiple domains, including military operations (Engel, 1954; Samz, 1972; Taylor and Brown, 1976; Taylor, 1979; Bracken, 1995; Fricker, 1998; Turkes, 2000; Lucas and Turkes, 2004; Hung, 2005; Schramm, 2012; Kress, et al., 2018; Stymfal, 2022; Cangiotti, et al., 2023), market competition dynamics (Shinichi, 1996; Fehlmann, 2008), ecological population dynamics (Shelley, et al., 2004), evolutionary psychology (Dominic and MacKay, 2015), and cybersecurity confrontations (Liu, et al., 2013). These systems employ first-order differential equations to characterize temporal force strength evolution in adversarial engagements. Analytical solutions derived under specified initial conditions facilitate comprehensive understanding of force contrast dynamics through explicit mathematical expressions.

While deterministic solutions under fixed initial conditions provide foundational insights, they may exhibit inherent limitations in modeling real-world scenarios characterized by stochastic fluctuations and emergent complexity. Traditional Lanchester formulations, despite their mathematical robustness, require enhanced adaptability to variable operational parameters for improved relevance in contemporary conflict analysis. This study introduces a novel methodological framework incorporating both initial and final state conditions, with the latter determined through variation principles of least action, thereby mitigating solution rigidity inherent in classical approaches.

The principle of least action, formally established as a fundamental principle in theoretical physics (Terekhovich, 2015), postulates that physical systems evolve along the extremal action function trajectories, typically minimizing the action integral. This principle operationalizes concepts of systemic efficiency and optimization within mechanical frameworks. Through variation calculus, prescribed boundary conditions yield Euler-Lagrange equations governing system dynamics (Chaichian and Demichev, 2001; Hanc, et al., 2003). Contemporary applications extend beyond classical physics to encompass electronic information systems (Huang, et al., 2024), computational image processing (Malashin, 2019), and behavioral psychology (Gibson, 1900), where it provides a framework for elucidating optimal resource allocation strategies under constrained conditions.

This study posits that combat strength evolution follows least-action trajectories, with force deployment patterns minimizing action functionals. The methodological framework is demonstrated through application to second-order temporal derivative version of Lanchester's square law, where the action functional is constructed through inverse variation principles to recover the governing Euler-Lagrange equation. This constitutes a least-action-enhanced Lanchester's square law model (hereinafter referred to as the Least Action Model, LAM), which enables predictive modeling of force attrition dynamics extending beyond the initial engagement phase when augmented with partial observational data. Empirical validation through historical analysis of the Battle of Kursk and Iwo Jima Campaign demonstrates the model's capability to generate predictive estimates for subsequent phases using limited initial engagement data. Comparative analysis with classical Lanchester's square law predictions reveals that the proposed variation framework demonstrates a statistically significant improvement in predictive accuracy for force strength trajectories.

This research aims to enhance the understanding of combat dynamics and establish innovative methodological frameworks for optimizing the strategic deployment of materials and personnel in relevant operational fields. The research results have the potential to clarify the fundamental mechanisms underlying conflict resolution processes and provide data - driven decision - making protocols for optimizing resource allocation.

## 2　The Lanchester's square law and its extension

Lanchester's square law (LSL) provides a mathematical framework for analyzing attrition dynamics in modern direct-fire combat scenarios. The model assumes mutually exposed opposing forces where each combat unit maintains direct targeting capability against adversaries. Within this paradigm, the attrition rate sustained by either belligerent demonstrates proportional dependence on both the numerical strength and combined combat effectiveness (encompassing firepower and survivability parameters) of the opposing force. For a closed system comprising Red and Blue forces engaged in combat without reinforcement capabilities, the fundamental governing equations of attrition can be expressed as

$$\begin{cases} \dot{q}_r = -k_b q_b, \\ \dot{q}_b = -k_r q_r \end{cases}, \tag{1}$$

where $q_r(t)$ and $q_b(t)$ represent the time-dependent quantities of red and blue combatants respectively, and $\dot{q}(t)$ signifies the first-order time derivative of $q(t)$.

The parameters $k_r$ and $k_b$ correspond to the combat power coefficients characterizing the operational effectiveness of respective forces. These coefficients may be determined through empirical parameter estimation derived from historical engagement data or preliminary battlefield observations.

Temporal differentiation of both members of the equation produces a system of second-order differential equations

$$\begin{cases} \ddot{q}_r(t) = k_r k_b q_r(t) \\ \ddot{q}_b(t) = k_r k_b q_b(t) \end{cases}. \tag{2}$$

The general solution structure can be expressed as

$$\begin{cases} q_r(t) = A e^{\omega t} + B e^{-\omega t} \\ q_b(t) = C e^{\omega t} + D e^{-\omega t} \end{cases}, \tag{3}$$

where $\omega = \sqrt{k_r k_b}$. The undetermined coefficients $A$, $B$, $C$, and $D$ represent integration constants that can be uniquely determined through application of four integral constraint conditions.

The most prevalent and empirically validated prescription involves the initial values of $q(t)$ and $\dot{q}(t)$ for both combatant factions. Furthermore, these parameters correspond to $q_r(t_0)$, $q_b(t_0)$, and

$$\begin{cases} \dot{q}_r(t_0) = -k_b q_b(t_0) \\ \dot{q}_b(t_0) = -k_r q_r(t_0) \end{cases}, \tag{4}$$

from which the particular solution can be derived

$$\begin{cases} q_r(t) = q_r(t_0) \cosh\left(\sqrt{k_r k_b}\, t\right) - \sqrt{\dfrac{k_b}{k_r}}\, q_b(t_0) \sinh\left(\sqrt{k_r k_b}\, t\right) \\ q_b(t) = q_b(t_0) \cosh\left(\sqrt{k_r k_b}\, t\right) - \sqrt{\dfrac{k_r}{k_b}}\, q_r(t_0) \sinh\left(\sqrt{k_r k_b}\, t\right) \end{cases}, \tag{5}$$

where $t_0$ represents the combat initiation temporal parameter. It is readily verifiable that Equation (5) constitutes the exact solution set for the LSL (1) under specified initial conditions $q_r(t_0)$ and $q_b(t_0)$. Given parameter values $k_r$ and $k_b$ coupled with initial force strengths, the temporal evolution of combatant quantities becomes uniquely determined through these solutions. Although this mechanistic determinism provides theoretical rigor to Lanchester-type models, it may inherently impose structural constraints that potentially limit their capacity to comprehensively model complex attrition dynamics.

In this study, we present an enhanced flexibility method for solving second-order differential equation (2) through the implementation of differentiated integral conditions. These boundary constraints are formulated by specifying both initial value $q(t_0)$ and terminal value $q(\tau)$ for the generalized coordinate $q(t)$ on opposing combatant sides. The terminal condition is derived through rigorous application of the least action principle, which we posit maintains validity in military conflict modeling. It should be noted that solutions generated by this methodology do not strictly conform to solutions of the classical LSL (1) with prescribed initial conditions $q_r(t_0)$ and $q_b(t_0)$, but rather constitute a generalized extension that accommodates variation boundary parameters.

Given the known boundary values of generalized coordinates at the initial time $t_0$ and terminal moment $\tau$, it is possible to derive four independent equations from which the four integration constants in equation (3) can be determined by solving these equations

$$\begin{cases} A = \dfrac{e^{\omega t_0} q_r(t_0) - e^{\omega \tau} q_r(\tau)}{e^{2\omega t_0} - e^{2\omega \tau}} \\[2pt] B = \dfrac{e^{\omega(t_0+\tau)}\left[q_r(\tau)e^{\omega t_0} - q_r(t_0)e^{\omega \tau}\right]}{e^{2\omega t_0} - e^{2\omega \tau}} \\[2pt] C = \dfrac{e^{\omega t_0} q_b(t_0) - e^{\omega \tau} q_b(\tau)}{e^{2\omega t_0} - e^{2\omega \tau}} \\[2pt] D = \dfrac{e^{\omega(t_0+\tau)}\left[q_b(\tau)e^{\omega t_0} - q_b(t_0)e^{\omega \tau}\right]}{e^{2\omega t_0} - e^{2\omega \tau}} \end{cases} \quad (6)$$

Therefore, the solution to the extended Lanchester's square system is

$$\begin{cases} q_r(t) = \operatorname{csch}(\omega(t_0-\tau))\left[q_r(t_0)\sinh(\omega(t-\tau)) - q_r(\tau)\sinh(\omega(t-t_0))\right] \\ q_b(t) = \operatorname{csch}(\omega(t_0-\tau))\left[q_b(t_0)\sinh(\omega(t-\tau)) - q_b(\tau)\sinh(\omega(t-t_0))\right] \end{cases}, \quad (7)$$

With $t \in [t_0, \tau]$.

## 3 The principle of least action

The principle of least action posits that among all conceivable trajectories connecting an initial state $A$ at time $t_1$ to a final state $B$ at time $t_2$, the physical realization of a system's motion at $t_1 < t < t_2$ corresponds to the path that minimizes the action functional

$$\mathcal{S}[q] = \int_{t_1}^{t_2} \mathcal{L}(q(t), \dot{q}(t), t)\, dt, \quad (8)$$

where the integrand $\mathcal{L}(q, \dot{q}, t)$, designated as the Lagrangian function, exhibits explicit dependence on the generalized coordinates $q(t)$ and their temporal derivatives $\dot{q}(t)$. This functional dependence fundamentally governs the system's dynamical behavior.

Under infinitesimal variations $\delta q$ of the coordinate $q(t)$ with fixed endpoints $\delta q(t_1) = \delta q(t_2) = 0$, the action functional $\mathcal{S}[q]$ undergoes corresponding variations $\delta \mathcal{S}[q]$ defined by $\delta \mathcal{S}[q] := \mathcal{S}[q(t) + \delta q(t)] - \mathcal{S}[q(t)]$. Through

variation calculus, this minimization principle $\delta \mathcal{S}[q]=0$ yields the Euler-Lagrange equation

$$\frac{\partial \mathcal{L}}{\partial q_i} - \frac{d}{dt}\left(\frac{\partial \mathcal{L}}{\partial \dot{q}_i}\right) = 0. \tag{9}$$

This differential equation in the temporal domain constitutes the fundamental equation of motion for classical dynamical systems.

While the theoretical framework suggests that all differential equation-governed systems should admit formulation through this variation principle, the principal challenge resides in the rigorous construction of appropriate action functionals. Physical systems typically derive their Lagrangian formulations through empirical phenomenology or symmetry principles. In classical mechanics, Newtonian dynamics (experimentally established through second law verification) permits derivation of the Lagrangian $\mathcal{L} = T - V$, where $T = \frac{1}{2}m\dot{q}^2$ represents the kinetic energy of a mass $m$ particle and $V$ its potential energy. This paradigmatic approach informs our methodology for analyzing military conflicts through Lanchester's laws (Hanc, et al., 2003). Focusing specifically on the LSL as our analytical foundation, we systematically develop a mechanical analogy for combat dynamics.

## 4  Attrition model from the least action principle in Lanchester's square law

Not all differential equations inherently possess an action formulation. As a first-order differential system, LSL presents such a scenario where direct derivation of the Lagrangian from Equation (1) through the Euler-Lagrange equation framework proves infeasible. Our approach consequently initiates from the second-order

differential equation

$$\begin{cases} \ddot{q}_r(t) = \omega_r^2 q_r(t) \\ \ddot{q}_b(t) = \omega_b^2 q_b(t) \end{cases}, \tag{10}$$

where the combat effectiveness coefficients $\omega_r(t)$ and $\omega_b(t)$ have been strategically assigned to opposing forces for generalized modeling. This formulation extends beyond classical Lanchester theory, with the corresponding Lagrangian expressed as

$$\mathcal{L} = \dot{q}_r^2(t) + \dot{q}_b^2(t) + \omega_r^2(t) q_r^2(t) + \omega_b^2(t) q_b^2(t). \tag{11}$$

Through substitution of Equation (11) and Equation (7) into Equation (8), we derive the action functional $\mathcal{S}$ as a parametric function of $q(\tau)$

$$\begin{aligned}
\mathcal{S} &= \int_{t_0}^{\tau} dt \mathcal{L}(q, \dot{q}, t) = \int_{t_0}^{\tau} dt (\dot{q}_r^2 + \dot{q}_b^2 + \omega_r^2 q_r^2 + \omega_b^2 q_b^2) \\
&= \frac{2\omega_r e^{2\omega_r(t_0+\tau)}}{(e^{2\omega_r t_0} - e^{2\omega_r \tau})^2} [4 q_r(t_0) q_r(\tau) \sinh(\omega_r(t_0-\tau)) - (q_r^2(t_0) + q_r^2(\tau)) \sinh(2\omega_r(t_0-\tau))] \\
&+ \frac{2\omega_b e^{2\omega_b(t_0+\tau)}}{(e^{2\omega_b t_0} - e^{2\omega_b \tau})^2} [4 q_b(t_0) q_b(\tau) \sinh(\omega_b(t_0-\tau)) - (q_b^2(t_0) + q_b^2(\tau)) \sinh(2\omega_b(t_0-\tau))] \\
&= \mathcal{S}_r + \mathcal{S}_b
\end{aligned} \tag{12}$$

The systemic action manifests as the superposition of bilateral combat interactions. This formulation enables battlefield analysts to extract meaningful operational insights through unilateral force strength metrics, eliminating the necessity for comprehensive dual-force data acquisition.

Under the principle of least action, we posit that the system evolves along optimal trajectories determined through variation calculus. Given initial conditions, this principle uniquely specifies subsequent state $q_r(\tau)$ and $q_b(\tau)$ through

stationary action requirements

$$\begin{cases} \dfrac{\partial \mathcal{S}_r}{\partial q_r(\tau)} = 0 \\ \dfrac{\partial \mathcal{S}_b}{\partial q_b(\tau)} = 0 \end{cases},$$

which yields

$$\begin{cases} q_r(\tau) = \dfrac{2e^{\omega_r(t_0+\tau)}}{e^{2\omega_r t_0} + e^{2\omega_r \tau}} q_r(t_0) = 2\dfrac{\sinh(\omega_r(t_0-\tau))}{\sinh(2\omega_r(t_0-\tau))} q_r(t_0) \\ q_b(\tau) = \dfrac{2e^{\omega_b(t_0+\tau)}}{e^{2\omega_b t_0} + e^{2\omega_b \tau}} q_b(t_0) = 2\dfrac{\sinh(\omega_b(t_0-\tau))}{\sinh(2\omega_b(t_0-\tau))} q_b(t_0) \end{cases}. \quad (13)$$

The proportionality constant $f(\omega) = \dfrac{2e^{\omega(t_0+\tau)}}{e^{2\omega t_0} + e^{2\omega \tau}}$ governing the relationship between $q(\tau)$ and $q(t_0)$ resides within the interval $(0, 1]$, quantitatively describing force attrition during engagement.

This model finds its combat application in battlefield analysis by dynamically predicting attrition sequences through real-time data integration, casting new light on operational patterns with predictive precision. Given consecutive force strength data $q(t_1)$ and $q(t_2)$, the engagement parameter $\omega(t_1 \to t_2)$ can be determined via Equation (13)

$$\begin{cases} \omega_r(t_1 \to t_2) = \dfrac{1}{t_2-t_1} \ln\left(\dfrac{q_r(t_1) + \sqrt{q_r^2(t_1) - q_r^2(t_2)}}{q_r(t_2)}\right) \\ \omega_b(t_1 \to t_2) = \dfrac{1}{t_2-t_1} \ln\left(\dfrac{q_b(t_1) + \sqrt{q_b^2(t_1) - q_b^2(t_2)}}{q_b(t_2)}\right) \end{cases}. \quad (14)$$

Subsequently, we employ Equation (13) with the parameter $\omega(t_2 \to t_3) = \omega(t_1 \to t_2) + \delta\omega$ and the initial condition $q(t_2)$ to iteratively forecast the temporal progression of troop strength data. Through this computational procedure, the third temporal data point $q(t_3)$ is predicted, yielding the following

expression for the anticipated strength at time $t_3$

$$\begin{cases} q_r(t_3) = \dfrac{2\sinh\left[\left(\omega_r(t_1 \to t_2) + \delta\omega_r\right)(t_3 - t_2)\right]}{\sinh\left[2(\omega_r(t_1 \to t_2) + \delta\omega_r)(t_3 - t_2)\right]} q_r(t_2) \\ q_b(t_3) = \dfrac{2\sinh\left[\left(\omega_b(t_1 \to t_2) + \delta\omega_b\right)(t_3 - t_2)\right]}{\sinh\left[2(\omega_b(t_1 \to t_2) + \delta\omega_b)(t_3 - t_2)\right]} q_r(t_2) \end{cases}. \quad (15)$$

Under infinitesimal time increment assumptions $t_2 - t_1 = t_3 - t_2 = \delta t$, where $\delta\omega$ represents a perturbative quantity, second-order Taylor expansion in terms of $\delta\omega$ and $\delta t$ yields

$$\begin{cases} q_r(t_3) = \dfrac{q_r(t_2)}{q_r(t_1)} q_r(t_2) - q_r(t_2)\delta t\,\text{sech}\left[\omega_r(t_1 \to t_2)\delta t\right]\tanh\left[\omega_r(t_1 \to t_2)\delta t\right]\delta\omega_r \\ +\dfrac{1}{4}q_r(t_2)\delta t^2 \text{sech}^3\left[\omega_r(t_1 \to t_2)\delta t\right]\left(\cosh\left[\omega_r(t_1 \to t_2)\delta t\right] - 3\right)\delta\omega_r^2 \\ q_b(t_3) = \dfrac{q_b(t_2)}{q_b(t_1)} q_b(t_2) - q_b(t_2)\delta t\,\text{sech}\left[\omega_b(t_1 \to t_2)\delta t\right]\tanh\left[\omega_b(t_1 \to t_2)\delta t\right]\delta\omega_b \\ +\dfrac{1}{4}q_b(t_2)\delta t^2 \text{sech}^3\left[\omega_b(t_1 \to t_2)\delta t\right]\left(\cosh\left[\omega_r(t_1 \to t_2)\delta t\right] - 3\right)\delta\omega_b^2 \end{cases}$$

(16)

## 5 Analytical verification of combat dynamics via the battles of Iwo Jima and Kursk

To validate the LAM, this study applies it to analyze two historically significant World War II engagements: the Battles of Iwo Jima and Kursk. Engel's (1954) seminal verification of LSL established methodological precedent through quantitative analysis of daily U.S. troop deployment data from Iwo Jima. Subsequent confirmation by Samz (1972) using alternative combat records reinforced the model's robustness. Contemporary re-evaluation by Stymfal (2022) achieved exceptional congruence ($R^2$=0.9937) for the square law formulation, while demonstrating comparable adequacy for linear ($R^2$=0.9027) and logarithmic ($R^2$=0.9414) Lanchester models.

Conversely, the Eastern Front's Battle of Kursk presents analytical complexities that challenge conventional Lanchester frameworks. Turkes (2000) and Lucas (2004) systematically demonstrated the inadequacy of basic Lanchester laws through daily casualty analysis, with phase-segmented and force-weighted modeling attempts yielding statistically insignificant correlations. The observed attrition dynamics revealed time-variant coefficients, suggesting operational factors including terrain utilization and tactical adaptations that exceed the original model's deterministic parameters.

This investigation employs primary historical records from Morehouse (1946) documenting 36-day U.S. force strength variations during Iwo Jima, and CAA (1998) archival data detailing two-week engagements between Soviet and German frontline combatants at Kursk. Both datasets exclude support personnel while incorporating reinforcement contingents, with continuous daily records providing temporal resolutions for analysis.

## 5.1 Analysis approach

Contrary to conventional global fitting approaches employed in prior research, our methodology adopts a sequential analytical framework mirroring real-time combat data processing. The attrition data undergoes daily collection and modeling through three operational phases: systematic acquisition of friendly force attrition metrics, parametric estimation via model fitting, and subsequent force strength prediction. The parameter $\omega(t)$ is numerically determined through nonlinear regression analysis of Equation (13) against observational data, followed by predictive computation using Equation (16) for subsequent temporal intervals. The perturbative quantity $\delta\omega$ is set to $\delta\omega=0$ and to be operationalized as the mean differential of $\omega$ across all acquired data points respectively.

Three distinct temporal fitting regimes are implemented for parameter estimation: (i) cumulative data assimilation up to day $t$, (ii) two-day moving window regression, and (iii) three-day retrospective analysis. Comparative evaluation of predictive outcomes across these temporal regimes enables quantitative assessment of model stability and temporal sensitivity.

A parallel analytical framework was established using the LSL as a control system, maintaining identical procedural parameters and temporal sequencing. This comparative approach facilitates rigorous validation of the LAM, notwithstanding documented limitations of LSL in Kursk Campaign analysis (Lucas and Turkes, 2004). The derivation of LAM from fundamental LSL principles necessitates such controlled comparison for theoretical verification.

Quantitative evaluation of predictive performance employs three rigorously defined metrics: daily relative error (DRE), mean relative error (MRE), and coefficient of determination ($R^2$). The DRE provides localized measurement of daily prediction accuracy, while the latter two metrics enable comprehensive global assessment through complementary statistical perspectives. This dual-metric validation framework effectively discriminates between truly predictive models and those benefiting from dataset-specific coincidences.

The DRE for temporal interval $i$

$$DRE_i = \frac{|\hat{q}_i - q_i|}{q_i} \times 100\% \tag{17}$$

quantifies the relative deviation between model-predicted strength values $\hat{q}_i$ and empirically observed strength data $q_i$ for each daily observation $i$. The MRE represents the arithmetic mean of daily relative error calculated throughout the entire military engagement period.

The coefficient of determination ($R^2$) for a single side is expressed as

$$R^2 = 1 - \frac{\sum_{i=1}^{n}(\Delta q_i - \Delta \hat{q}_i)^2}{\sum_{i=1}^{n}(\Delta q_i - \overline{\Delta q})^2}, \tag{18}$$

where $\Delta q_i = q_i - q_{i+1}$ represents the actual force losses on day $i$, and $\overline{\Delta q}$ denotes the mean daily force losses throughout the engagement period. The term $\Delta \hat{q}_i = q_i - \hat{q}_{i+1}$ corresponds to the model-predicted force losses on day $i$. Higher $R^2$ values indicate superior predictive accuracy, with optimal performance achieving $R^2=1$. Under this formulation, negative $R^2$ values may occur, signifying that the model demonstrates inferior predictive performance compared to estimations derived from the mean daily loss value.

## 5.2 Results

Figures 1 and 2 illustrate the LAM and LSL predicting daily force dynamics of Soviet and German forces respectively during the Battle of Kursk, while Figure 3 demonstrates comparable predictions for U.S. forces in the Battle of Iwo Jima. The predictions incorporate reinforcement adjustments to initial force estimates.

Comparative analysis reveals that LAM's strength evolution curves exhibit superior alignment with historical records compared to LSL predictions. This observation is quantitatively supported by Figures 4-6, which present daily relative errors for Soviet and German forces at Kursk, and U.S. forces at Iwo Jima respectively. The analysis reveals that all DREs predicted by the LAM remain below 4% across both battles, whereas certain predictions from the LSL exhibit the errors exceeding 5%. A striking disparity emerges in the DREs of German forces during the Battle of Kursk. The comparative analysis presented in Figure 5 demonstrates that LAM predictions maintain consistently below 0.5% error, while corresponding LSL estimates during initial combat phases surpass 7%.

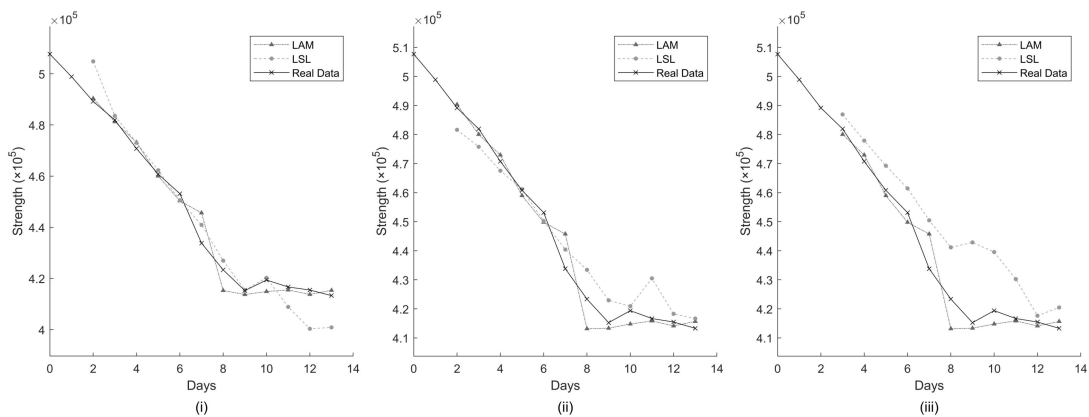

Figure 1. Comparative force predictions for Soviet forces in the Battle of Kursk across scenarios (i)-(iii), contrasting least action model (LAM) and Lanchester's square law (LSL) estimates with historical records.

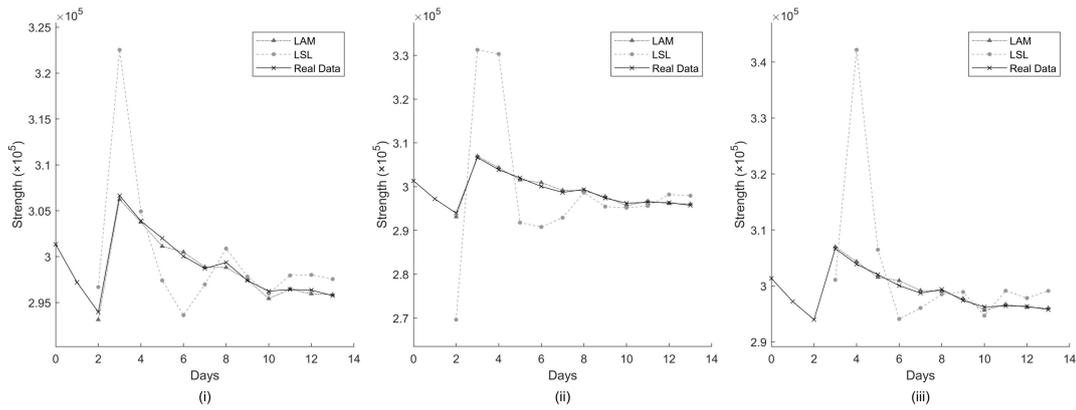

Figure 2. Comparative force predictions for German forces in the Battle of Kursk across scenarios (i)-(iii), contrasting least action model (LAM) and Lanchester's square law (LSL) estimates with historical records.

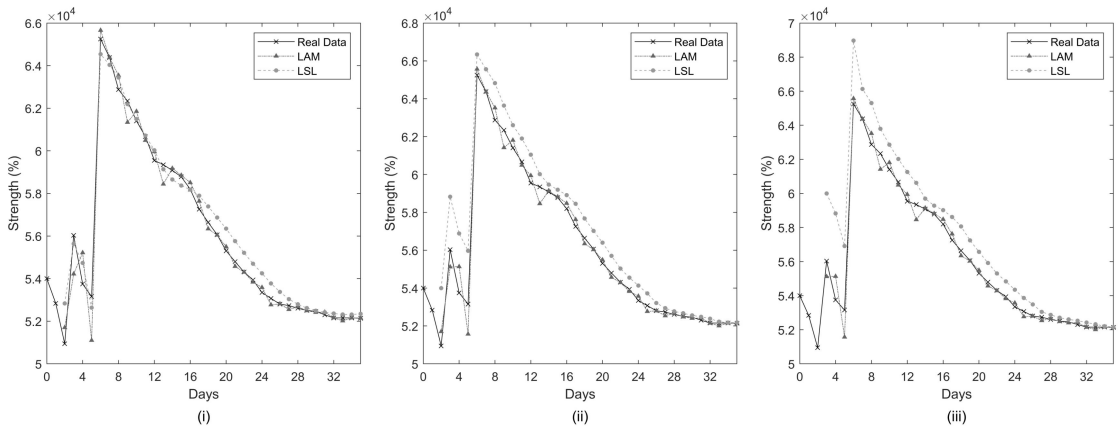

Figure 3. Comparative force predictions for U.S. force dynamics in the Battle of Iwo Jima across scenarios (i)-(iii), contrasting least action model (LAM) and Lanchester's square law (LSL) estimates with historical records.

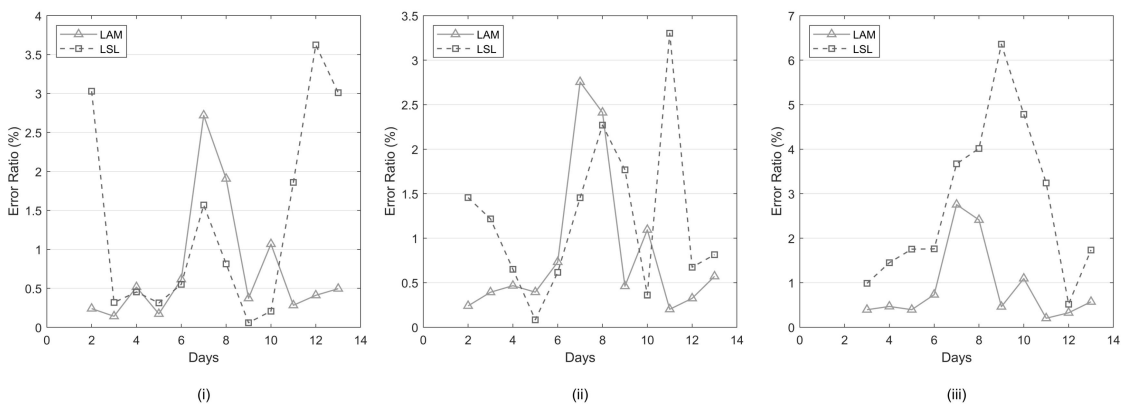

Figure 4. Daily relative error predictions for Soviet forces in the Battle of Kursk, contrasting least action model (LAM) and Lanchester's square law

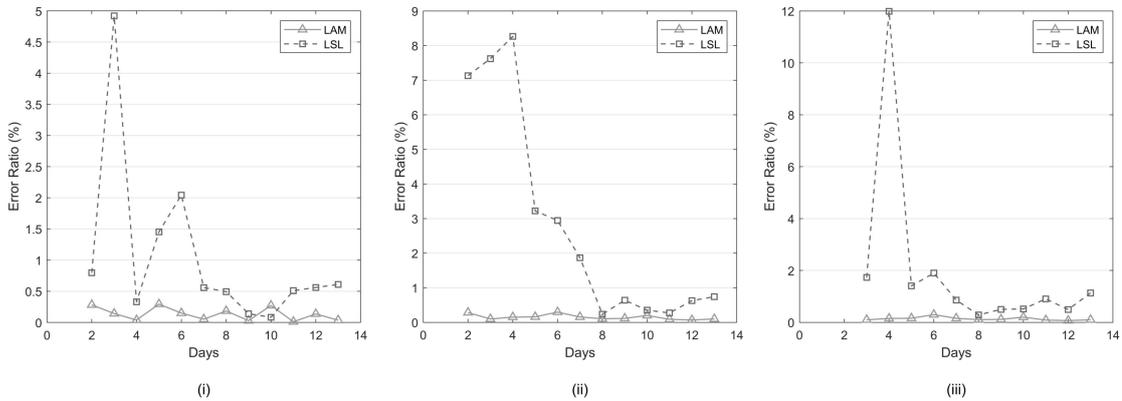

Figure 5. Daily relative error predictions for German force in the Battle of Kursk, contrasting least action model (LAM) and Lanchester's square law (LSL) performance across three scenarios.

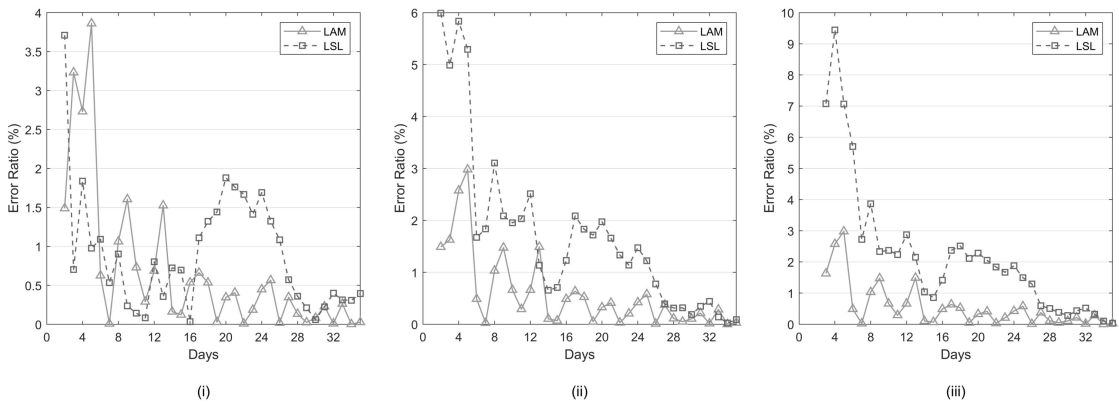

Figure 6. Daily relative error predictions for U.S. forces in the Battle of Iwo Jima, contrasting least action model (LAM) and Lanchester's square law (LSL) performance across three scenarios.

Table 1 summarizes comparative evaluation metrics (Mean Relative Error [MRE], Coefficient of Determination [$R^2$]) for both analytical engagements, quantitatively assessing the predictive capabilities of both LAM and LSL frameworks. The analysis reveals LAM's statistically superior performance across both operational theaters, with MRE metrics demonstrating consistent outperformance over LSL counterparts. Notably, while LSL exhibits negative $R^2$ values in Kursk engagement predictions - indicative of predictive performance inferior to simple mean-value benchmarks - LAM maintains robust positive explanatory power across all combat scenarios. While both models exhibit comparable determination coefficients ($R^2$) in Iwo Jima scenario (i), the LAM manifests enhanced robustness across operational parameter variations through scenarios (ii)-(iii).

Table 1. Comparative Evaluation Metrics for Predictive Models Across Historical Engagements.

| Data | Parameter fitting scenario | MRE | | $R^2$ | |
|---|---|---|---|---|---|
| | | LAM | LSL | LAM | LSL |
| Kursk:Soviet | (i) | 0.7448% | 1.3503% | 0.3698 | -0.9292 |
| | (ii) | 0.8369% | 1.2634% | 0.2467 | -0.1635 |
| | (iii) | 0.8912% | 2.8413% | 0.2410 | -0.0973 |
| Kursk:German | (i) | 0.1351% | 1.0930% | 0.9845 | -0.6058 |
| | (ii) | 0.1503% | 3.0246% | 0.9848 | -9.9595 |
| | (iii) | 0.1383% | 2.0582% | 0.9877 | -4.3556 |
| Iwo Jima:USA | (i) | 0.6760% | 0.8926% | 0.9283 | 0.9304 |
| | (ii) | 0.5826% | 1.7194% | 0.9532 | 0.7438 |
| | (iii) | 0.5552% | 2.2361% | 0.9553 | 0.5748 |

## 6  Summary and conclusions

The conventional Lanchester equations demonstrate inherent limitations in characterizing force attrition dynamics within modern military engagements, particularly manifesting critical inadequacies when modeling nonlinear attrition patterns and multidimensional battlefield environments. To resolve these theoretical constraints, this study proposes an enhanced analytical framework grounded in variation principles, specifically implementing least action methodology to improve predictive fidelity while establishing an adaptive theoretical architecture for comprehensive combat analysis. Using the LSL as a representative case, we formulate a force attrition model through variation calculus and empirically validate its performance through detailed analysis of force progression data from the WWII Battle of Kursk and Battle of Iwo Jima.

Empirical validation reveals three principal findings: First, the least action-enhanced Lanchester model achieves superior predictive accuracy compared to conventional LSL implementations across multiple historical combat scenarios. The LAM maintains daily relative prediction errors under 4% across all engagements, significantly surpassing LSL performance that periodically exceeds 5% error margins. Second, the model demonstrates exceptional precision in modeling German force attrition at Kursk, sustaining sub-0.5% prediction errors compared to LSL's 7% discrepancies during initial combat phases. Third, while both models exhibit comparable determination coefficients ($R^2$) in Iwo Jima scenario (i), the LAM manifests enhanced robustness across operational parameter variations through scenarios (ii)-(iii).

This theoretical integration of variation mechanics with combat modeling enables three critical advancements: 1) Dynamic adaptation to force reinforcement patterns through energy potential formulations; 2) Precise characterization of nonlinear attrition dynamics via action functional minimization; 3) Consistent maintenance of positive $R^2$ values across theaters, contrasting with LSL's frequent negative determinations that fail basic mean-value benchmarks. Particularly noteworthy is LAM's sustained explanatory power ($R^2>0.95$) in high-intensity attrition environments like the German Kursk campaign, where conventional models exhibit catastrophic predictive failures.

Future research should prioritize three investigative axes: 1) Extension to asymmetric warfare paradigms incorporating technological disparity coefficients; 2) Temporal-spatial combat variable integration through Hamiltonian phase-space formulations; 3) Development of real-time predictive architectures for network-centric warfare applications. The demonstrated efficacy in modeling 20th-century conventional engagements suggests substantial potential for contemporary conflict analysis, contingent upon parameter space expansion to encompass cyber-electronic warfare dimensions and advanced sensor network inputs.